\documentstyle[preprint,epsfig,aps]{revtex}

\begin{document}
\draft
\tighten
\title{Two-neutrino double beta decay within Fully-Renormalized QRPA:
Effect of the restoration of the Ikeda sum rule}
%The influence of the restoration of the Ikeda sum rule within Fully-Renormalized QRPA
%on the two-neutrino double beta decay amplitude}
\author{L. Pacearescu$^1$, V. Rodin$^1$, F. \v Simkovic$^{1,2}$ and Amand Faessler$^1$}
\address{$^1$
Institut f\"ur Theoretische Physik der Universit\"at T\"ubingen,\\
Auf der Morgenstelle 14, D-72076 T\"ubingen, Germany}
\address{$^2$Department of Nuclear Physics, Comenius University,
Bratislava, Slovakia}
\date{\today}
\maketitle
 \begin{abstract}
The recently proposed Fully-Renormalized QRPA (FR-QRPA), which fullfils the Ikeda sum rule (ISR) exactly,
%formulated recently
is applied to the two-neutrino double beta decay of
$^{76}$Ge, $^{82}$Se, $^{100}$Mo, $^{116}$Cd, $^{128}$Te  and $^{130}$Xe.
%Rather small model bases are used to get convergence in the model.
The results obtained are compared with those of other approaches, 
standard QRPA and self-consistent QRPA (SCQRPA).
The similarities and the differences among the methods are discussed.
The influence of the restoration of the Ikeda sum rule on the
$2\nu\beta\beta$-decay amplitude is analyzed.
\end{abstract}

\section{Introduction}

Observation of the neutrinoless double beta decay ($0\nu\beta\beta$-decay), violating
the total lepton number by two units, would give unambiguous evidence
for new physics beyond the Standard Model~\cite{fae98,hax84,Vog}. For instance,
at least one of the neutrinos would have to be a Majorana particle with non-zero mass~\cite{sch82}.
The current experimental
upper limits on the $0\nu\beta\beta$-decay half-life impose stringent
constraints, e.g.,  on  the parameters
of Grand Unification and super-symmetric extensions
of the Standard Model.

Rates of the $2\nu\beta\beta$-decay, which is a second
order process allowed within the Standard Model, can be calculated
within the same nuclear structure models. Thus, the
results of the nuclear structure calculations can be directly compared
with the corresponding experimental data available for a number of nuclei~\cite{exp}.
Such a comparison provides a very useful test of the models.

The Quasiparticle Random Phase Approximation (QRPA)~\cite{book}
has been successfully exploited in nuclear physics to describe properties of the excited states
of open-shell nuclei and to calculate intensities of
various nuclear reactions, including the double beta decay (see reviews \cite{fae98}).

It was shown that the experimental data on the $2\nu\beta\beta$-decay rates can
be reproduced in QRPA calculations with a sufficiently large strength of the particle-particle interaction~\cite{vog86}.
But the proximity of the value to the point of the QRPA collapse questions the
reliability of the results. It is known that
QRPA collapse occurs due to the use of the quasi-boson approximation (QBA) which
violates the Pauli exclusion principle (PEP) and generates too many ground state correlations.

Renormalized QRPA (RQRPA) was formulated in Refs.~\cite{rqrpa} to restore PEP in an approximate way.
The main goal of the method is to use a self-consistent iteration of the QRPA equation
with taking into account quasiparticle occupation numbers in the QRPA ground state.
That leads to
a modification of the commutation relations for bifermionic operators as compared to the ordinary quasiboson approximation (QBA).
At the same time so-called scattering terms (describing transitions of the
quasiparticles) are neglected in the Hamiltonian and in the phonon operators.
The RQRPA does not collapse for physical values of the particle-particle interaction strength
and has been extensively used to calculate the intensities of the double beta decay \cite{fae98,FKSS97,Toi97}.
It has been also shown that the RQRPA provides better agreement with the exact solution of
the many-body problem within schematic models, even beyond the critical point
of the standard QRPA (see, e.g. \cite{schm} and references therein).

The self-consistent RQRPA (SCQRPA) is a more complex version of RQRPA to describe
the strongly correlated Fermi systems. Within this method one goes a step further beyond the
RQRPA. In the SCQRPA at the same time the quasiparticle mean field is changed by minimizing the energy
and fixing the number of particles in the correlated ground state of RQRPA instead of the uncorrelated
one of BCS as is done in the other versions of the RQRPA. In this way SCQRPA partially overcomes the
inconsistency between RQRPA and the BCS approach and is closer to a fully variational theory.

Nevertheless, the main drawback of the modern versions of RQRPA and SCQRPA is the violation
of the model-independent Ikeda sum rule (ISR)~\cite{Toi97,Sto01,Bob00}.
A modification of the phonon operator by including scattering terms
is needed in order to restore the ISR within RQRPA. The fully-Renormalized QRPA (FR-QRPA) was formulated in Ref.~\cite{Rod02}
for even-even nuclei in such a way that it complies with restrictions imposed by the commutativity of the phonon
creation operator with the total particle number operator.
It was shown analytically that the Ikeda sum rule is fulfilled within the FR-QRPA~\cite{Rod02}.
Also FR-QRPA is free from the spurious low-energy solutions which would be generated by the scattering terms
considered as additional degrees of freedom as suggested in \cite{Rad98}.

The aim of the paper is twofold. First, we would like to describe the FR-QRPA equations in more details (as compared with the original paper~\cite{Rod02})
for a simple case of a Hamiltonian with the separable residual interaction in both particle-hole and particle-particle channels. Second, the first numerical application
of FR-QRPA is given to calculate $2\nu\beta\beta$-decay intensities and  relevant quantities. 
So far, the full convergence of the FR-QRPA solution
has been obtained only for a rather small model space. Nevertheless  a comparison of the results obtained within FR-QRPA and SCQRPA can be provided.

\section{Basic relationships of the Fully-Renormalized QRPA}

Within RPA an excited nuclear state, with angular momentum $J$ and projection $M$,
is created by applying the phonon operator $Q^{\dagger}_{JM}$
to the vacuum state $|0^+_{RPA}\rangle$ of the initial, even-even, nucleus:
%with the same nucleon number $A$:
\begin{equation}
|JM  \rangle = Q^{\dagger}_{JM }|0^+_{RPA}\rangle
 \qquad \mbox{with} \qquad
Q_{JM }|0^+_{RPA}\rangle=0.
%\label{eq:19}
\end{equation}

As was shown in Ref.~\cite{Rod02}, the most appropriate way is to
write down the phonon structure in terms of the particle creation and 
annihilation operators.
That allows to fulfill the important principle of the
commutativity of $Q^{\dagger}_{JM}$ with the total particle number operator $\hat A =\hat N + \hat Z$.
%$\sum\limits^{}_{p , m_p}c^\dagger_{p m_p}{{c}}_{p m_p}$.
The phonon operator has the following structure:
\begin{equation}
Q^{\dagger}_{JM } = \sum\limits_{pn}
 \left [x_{(pn, J )} C^\dagger(pn, JM)
-y_{(pn, J )}\tilde{C}(pn, JM)\right ],
\label{Qc}
\end{equation}
with $C^\dagger(pn, JM)=\left[c^\dagger_{p}{\tilde{c}}_{n}\right]_{JM}$ and
 $\tilde C(pn, JM)=(-)^{J-M}C(pn, J\,-M)$,
where $c^{+}_{\tau m_{\tau}}$ ($c^{}_{\tau m_\tau}$) denotes the particle creation 
(annihilation) operator for protons and neutrons ($\tau=p,n$).
%The tilde symbol indicates the  time-reversal operation, e.g.$\tilde{a}_{p {m}_{p}}$ =$(-1)^{j_{p} - m_{p}}a^{}_{p -m_{p}}$.

Going into the quasiparticle representation, the quasiparticle 
creation and annihilation operators
$a^{+}_{\tau m_{\tau}}$ and $a^{}_{\tau m_{\tau}}, (\tau=p,n)$ can be defined by the Bogolyubov transformation
\begin{equation}
\left( \matrix{ a^{+}_{\tau m_{\tau} } \cr
 {\tilde{a}}_{\tau  m_{\tau} }
}\right) = \left( \matrix{
u_{\tau} & v_{\tau} \cr
-v_{\tau} & u_{\tau}
}\right)
\left( \matrix{ c^{+}_{\tau m_{\tau}} \cr
{\tilde{c}}_{\tau m_{\tau}}
}\right),
\label{uv}
\end{equation}
that leads to the following expression for the phonon operator $Q^{\dagger}_{JM}$:
\begin{eqnarray}
&Q^{\dagger}_{JM } = \sum\limits_{{p}{n}}
 \left [ X_{({p}{n}, J )} \bar A^\dagger({p}{n}, JM)
- Y_{({p}{n}, J )}\tilde{\bar A\,}({p}{n}, JM)\right ],
\label{Qa1} \\
&\nonumber\\
& \bar A^\dagger= A^\dagger+\left(u_{n}v_{n}B^\dagger-
u_{p}v_{p}\tilde B\right)\left/(v_{n}^2-v_{p}^2)\right. \nonumber\\
& A^\dagger(pn, JM)=\left[a^\dagger_{p} a^\dagger_{n}\right]_{JM}\ ; \ \
B^\dagger(pn, JM)=\left[a^\dagger_{p}{\tilde{a}}_{n}\right]_{JM}\nonumber
\end{eqnarray}
where $X=u_{p}v_{n}x-v_{p}u_{n}y,\ Y=u_{p}v_{n}y-v_{p}u_{n}x$.
The bifermionic operators $\bar A^\dagger,\bar A$ now being the basic building blocks of the FR-QRPA
automatically contain the quasiparticle scattering terms which, however, are not associated
with any additional degree of freedom. That means that there are no spurious low-lying solutions in the present theoretical scheme
which would be generated by the scattering terms considered as independent constituents of the phonon operator (as proposed in~\cite{Rad98}).

From this point we can follow the usual way to formulate the RQRPA~\cite{rqrpa},
substituting $A$ by $\bar A$ everywhere.
The forward- and backward-going free variational amplitudes X and Y satisfy the equation:
\begin{equation}
\left(
\begin{array}{cc}
{\cal A}&{\cal B}\\
{\cal B}&{\cal A}
\end{array}
\right)
\left(
\begin{array}{c}
X^m\\
Y^m
\end{array}
\right)
= {\cal E}_m
\left(
\begin{array}{cc}
{\cal U}&0\\
0&{\cal -U}
\end{array}
\right)
\left(
\begin{array}{c}
X^m\\
Y^m
\end{array}
\right),
\label{QRPA}
\end{equation}
where $m$ marks different roots of the QRPA equations for a
given $J^\pi$,
\begin{eqnarray}
{\cal A} &=& \langle 0^+_{RPA}| \left[ \bar A, \left[ H, \bar
A^\dagger \right ]\right]
|0^+_{RPA} \rangle, \nonumber \\
{\cal B} &=& - \langle 0^+_{RPA}| \left[ \bar A, \left[ H, \bar
A \right ]\right] |0^+_{RPA}
\rangle,
\label{eq:14}
\end{eqnarray}
and the renormalization matrix ${\cal U}_{pn}$ is
\begin{eqnarray}
{\cal U}_{pn}&=&\langle 0^+_{RPA}|\left[ \bar A(pn, JM),
\bar A^\dagger(p'n', JM)\right]|0^+_{RPA}\rangle =
\delta_{pp'}\delta_{nn'}
 {\cal D}_{pn}
.\label{Dm}
\end{eqnarray}

We use a rather simple, but realistic, Hamiltonian $H$ consisting of the quasiparticle mean field
$H_0$ and the residual separable particle--hole (ph) and particle--particle
(pp) interactions:
\begin{eqnarray}
& H = H_0 + H^{ph}_{int}+ H^{pp}_{int},\\
& H_0 = \sum\limits_{\tau=p,n} E_{\tau} a^\dagger_{\tau} a_{\tau},\\
& H_{int}^{ph} = ~\chi \sum\limits_{M}  (-1)^M ( \beta^-_{1M}\beta^+_{1-M}
+ \beta^+_{1-M} \beta^-_{1M}),\\
& H_{int}^{pp}= - \kappa \sum\limits_{M}   (-1)^M ( P^-_{1M}P^+_{1-M}
+ P^+_{1-M} P^-_{1M} ),
\end{eqnarray}
with $\beta_{1M}^-=-\hat J^{-1}
\sum\limits_{pn}\langle p\| \sigma\| n\rangle \left[c^\dagger_{p}{\tilde{c}}_{n}\right]_{1M}$,
$P_{1M}^-=\hat J^{-1}
\sum\limits_{pn}\langle p\| \sigma\| n\rangle \left[c^\dagger_{p}c^\dagger_{n}\right]_{1M}$ and $J=1$.

Taking into account the exact (fermionic) expressions for the commutators
in (\ref{eq:14}),(\ref{Dm}), one gets the following expressions for
the FR-QRPA matrices ${\cal A}$ and ${\cal B}$:
\begin{eqnarray}
{\cal A} &=& \left[(E_{p}+E_{n}) {\cal D}_{pn}-
2(E_{p}-E_{n})(u_{p}^{2}v_{p}^{2}+
u_{n}^{2}v_{n}^{2}){\cal R}_{pn}
%\frac{{\cal N}_{p}-{\cal N}_{n}}{v_{n}^{2}-v_{p}^{2}}
\right] \delta_{pp'} \delta_{nn'}\nonumber\\
&&+2\chi (u_{p}v_{n}u_{p'}v_{n'}+
v_{p}u_{n}v_{p'}u_{n'}) {\cal D}_{pn} {\cal D}_{p'n'}\nonumber\\
&&-2\kappa (u_{p}u_{n}u_{p'}u_{n'} {\bar{\bar {\cal D}}}_{pn} {\bar{\bar {\cal D}}}_{p'n'}+
v_{p}v_{n}v_{p'}v_{n'} {\bar {\cal D}}_{pn} {\bar {\cal D}}_{p'n'} ) \label{A},\\
\nonumber\\
{\cal B} &=& 2(E_{p}-E_{n}) u_{p}v_{p}u_{n}v_{n}
{\cal R}_{pn}\delta_{pp'}\delta_{nn'}\nonumber\\
&&+2\chi(u_{p}v_{n}v_{p'}u_{n'}+
v_{p}u_{n}v_{p'}u_{n'}){\cal D}_{pn}{\cal D}_{p'n'}\nonumber\\
&&+2\kappa(u_{p}u_{n}v_{p'}v_{n'}{\bar{\bar {\cal D}}}_{pn}
{\bar {\cal D}}_{p'n'}+
v_{p}v_{n}u_{p'}u_{n'}{\bar {\cal D}}_{pn}{\bar{\bar {\cal D}}}_{p'n'})
.\label{B}
\end{eqnarray}

The renormalization matrices ${\cal D},{\bar {\cal D}},{\bar{\bar {\cal D}}}$ entering (\ref{Dm}),(\ref{A}),(\ref{B})
can be represented in terms of the relative
quasiparticle occupation numbers ${\cal N}_{p}$ for the level $p$ in the RQRPA
vacuum:
\begin{eqnarray}
{\cal D}_{pn}& = &
%1+\left((u_{n}^2-v_{n}^2){\cal  N}_{n}-(u_{p}^2-v_{p}^2){\cal N}_{p}\right)\left/ (v_{n}^2-v_{p}^2)\right.
1-{\cal  N}_{n}-{\cal  N}_{p}+\left(1-v_{p}^2-v_{n}^2\right){\cal R}_{pn}
,\label{Dpn}\\
%{\bar {\cal D}}_{pn} &=& 1-2\left(u_{p}^{2}{\cal N}_{p}-u_{n}^{2}{\cal N}_{n}\right)\left/ (v_{n}^2-v_{p}^2)\right.\nonumber\\
{\bar {\cal D}}_{pn} &=& 1-{\cal  N}_{n}-{\cal  N}_{p}-\left(u_{p}^2+u_{n}^2\right){\cal R}_{pn}
,     \nonumber\\
%{\bar{\bar {\cal D}}}_{pn} &=& 1+2\left(v_{p}^{2}{\cal N}_{\tau}-v_{n}^{2}{\cal N}_{n}\right)\left/ (v_{n}^2-v_{p}^2)\right.
{\bar{\bar {\cal D}}}_{pn} &=& 1-{\cal  N}_{n}-{\cal  N}_{p}+\left(v_{p}^2+v_{n}^2\right){\cal R}_{pn}, \nonumber
\end{eqnarray}
with ${\cal R}_{pn}=\frac{{\cal N}_{p}-{\cal N}_{n}}{v_{n}^{2}-v_{p}^{2}}$.
In turn, the quasiparticle occupation numbers
\begin{equation}
{\cal N}_{p}=\hat j_{\tau}^{-2}\langle0^+_{RPA}|\sum\limits^{}_{m_{\tau}}
 a^\dagger_{\tau m_{\tau}}{{a}}_{\tau m_{\tau}}|0^+_{RPA}\rangle;~~~\tau=p,n.
%=-\hat j_p^{-1}\langle 0^+_{RPA}|B(pp,00)|0^+_{RPA}\rangle
\end{equation}
can be expressed in terms of the backgoing amplitudes $Y$
of the RQRPA solution (\ref{QRPA}) ~\cite{rqrpa}.
In the calculation we shall use the aproximate expression for ${\cal N}_{p}$ and ${\cal N}_{n}$:

\begin{eqnarray}
{\cal N}_{p} &\approx& \hat {j}_{p}^{-2}\sum_{n}
\left(\sum_{J,m} (2J+1)|Y^{m}_{pn,J}|^2  \right){\cal D}_{pn}\nonumber\\
{\cal N}_{n} &\approx& \hat {j}_{n}^{-2}\sum_{p}
\left(\sum_{J,m} (2J+1)|Y^{m}_{pn,J}|^2  \right){\cal D}_{pn}
\label{occup}
\end{eqnarray}
where $\hat j\equiv\sqrt{2j+1}$. 
In the present paper we consider only $J^{\pi}=1^{+}$ contribution to the sums 
in (\ref{occup}).
Along with
the modified SCQRPA and FR-QRPA equations for the chemical potential:
\begin{eqnarray}
\langle 0^+_{RPA}|\hat N |0^+_{RPA}\rangle&=&\sum\limits_{n}
\hat j_n^{2}\left(v_n^2+(u_n^2-v_n^2){\cal N}_{n}\right)=N
 %-\hat j_n^{-1}(u_n^2-v_n^2)\left<B(nn,00)\right>
,\nonumber\\
\langle 0^+_{RPA}|\hat Z|0^+_{RPA}\rangle&=&\sum\limits_{p}
\hat j_p^{2}\left(v_p^2+(u_p^2-v_p^2){\cal N}_{p}\right)=Z,
\label{lambdas}
\end{eqnarray}
a rather complicated set of equations (\ref{QRPA})-(\ref{lambdas}) has to be solved.

It is noteworthy that the renormalization matrices (\ref{Dpn}) become the same,
${\bar{\bar {\cal D}}}_{pn}={\bar {\cal D}}_{pn}={\cal D}_{pn}$,
in the limit ${\cal R}_{pn}=0$ and coinciding with the renormalization matrix of
the usual RQRPA (see,e.g.,~\cite{Toi97}). Thus, one can argue that the standard versions of RQRPA neglect
effectively the differences between the quasiparticle occupation numbers whereas SCQRPA and FR-QRPA
take the differences into account.

From now on we follow the usual way of solving RQRPA equations~\cite{rqrpa}. It is useful to introduce the notation:
\begin{equation}
\bar{X} = {\cal U}^{1/2} X, ~~~~~~\bar{Y} = {\cal U}^{1/2} Y,
\label{eq:15}
\end{equation}
\begin{equation}
\bar{\cal A} = {\cal U}^{-1/2} {\cal A} {\cal U}^{-1/2}, ~~~~
\bar{\cal B} = {\cal U}^{-1/2} {\cal B} {\cal U}^{-1/2}.
\label{eq:16}
\end{equation}
Then the amplitudes $\bar X$ and $\bar Y$ satisfy the equation of usual QRPA:
\begin{equation}
\left(
\begin{array}{cc}
\bar {\cal A}&\bar {\cal B}\\
\bar {\cal B}&\bar {\cal A}
\end{array}
\right)
\left(
\begin{array}{c}
\bar X^m\\
\bar Y^m
\end{array}
\right)
= {\cal E}_m
\left(
\begin{array}{cc}
1&0\\
0& -1
\end{array}
\right)
\left(
\begin{array}{c}
\bar X^m\\
\bar Y^m
\end{array}
\right).
%\label{dqrpa)
%\label{eq:13}
\end{equation}
%%%%%%%%%%%%%%%%%%%%%%%%%%%%%%%%%%%%%%%%%%%%%%
Solving the FR-QRPA  equations, one gets the fully
renormalized amplitudes $\bar{X}$, $\bar{Y}$
with the usual normalization and closure relations:
\begin{eqnarray}
\sum\limits_{pn}\bar{X}^m_{(pn, J )}\bar{X}^k_{(pn, J )}
-\bar{Y}^m_{(pn, J)}\bar{Y}^k_{(pn, J )}&=&\delta_{km},
\nonumber\\
\sum\limits_{m}\bar{X}^m_{(pn, J )}\bar{X}^m_{(p^{\phantom 1}_1n_1, J )}
-\bar{Y}^m_{(pn, J )}\bar{Y}^m_{(p^{\phantom 1}_1n_1, J )}&=&
\delta_{pp^{\phantom 1}_1}\delta_{nn_1},\nonumber\\
\sum\limits_{m}\bar{X}^m_{(pn, J )}\bar{Y}^m_{(p^{\phantom 1}_1n_1, J )}
-\bar{Y}^m_{(pn, J )}\bar{X}^m_{(p^{\phantom 1}_1n_1, J )}&=&0.
\label{closure}
\end{eqnarray}

It was shown analytically that the Ikeda sum rule is fulfilled within the FR-QRPA~\cite{Rod02}, in contrast to
the earlier versions of the RQRPA~\cite{rqrpa}.
The Ikeda sum rule states that the difference between the total
Gamow-Teller strengths $S^{(-)}$ and $S^{(+)}$ in the $\beta^-$ and $\beta^+$ channels, respectively, is
$3(N-Z)$~\cite{Ikeda}:
\begin{eqnarray}
&ISR=S^{(-)}-S^{(+)}=3(N-Z)\label{ISR},\\
&S^{(-)}=\sum\limits_{Mm} \left|\langle 1^{+}M,  m| \beta_{1M}^- | 0^+_{RPA}\rangle
\right|^2, ~~~~~ S^{(+)}=\sum\limits_{Mm'} \left|\langle 1^{+}M, m'| \beta_{JM}^+ | 0^+_{RPA}\rangle \right|^2.\nonumber\\
%&\nonumber
\end{eqnarray}
With the use of the closure conditions (\ref{closure}), the expressions for ${\cal D}_{pn}$ (\ref{Dpn}) and the chemical potentials
(\ref{lambdas}), one can show~\cite{Rod02} that
\begin{equation}
ISR=\sum_{pn} \left|\langle p\|q_{J} \| n\rangle\right|^2
(v_n^2-v_p^2) {\cal D}_{pn}=3(N-Z).
\end{equation}

The inverse half-life of the $2\nu\beta\beta$-decay  can be expressed as a product of  an accurately known
phase-space factor $G^{2\nu}$ and the second order Gamow-Teller transition matrix element $M^{2\nu}_{GT}$:
\begin{equation}
[ T^{2\nu}_{1/2}(0^+_{g.s.} \rightarrow 0^+_{g.s.}) ]^{-1} =
G^{2\nu} ~(g_A)^4~ | M^{2\nu}_{GT}|^2.
\label{halfl}
\end{equation}
The contribution from the two successive Fermi transitions is
safely neglected as they arise from isospin mixing effect \cite{hax84}.
The double Gamow-Teller matrix element $M^{2\nu}_{GT}$
for ground state to ground state $2\nu\beta\beta$-decay transition acquires the form
\begin{equation}
M^{2\nu}_{GT}=\sum_{{m_i m_f}}
\frac{\langle 0^+_f\parallel \beta^- \parallel 1^{+}_{m_f}\rangle
\langle 1^{+}_{m_f}|1^{+}_{m_i}\rangle
\langle 1^{+}_{m_i}\parallel \beta^- \parallel 0^+_i\rangle}
{(\omega^{m_f} + \omega^{m_i})/2}.
\label{betabeta}
\end{equation}
The sum extends over all $1^+$ states of the intermediate nucleus.
The index $i (f)$ indicates that the quasiparticles and the excited
states of the nucleus are defined with respect to the initial (final)
nuclear ground state $|0^+_i\rangle$ ($|0^+_f\rangle$). The overlap is necessary since
these intermediate states are not orthogonal to each other.
The two sets of intermediate nuclear states generated from the
initial and final ground states are not identical within the
considered approximation scheme. Therefore the overlap factor
of these states is introduced in the theory as follows:
\begin{equation}
\langle 1^{+}_{m_f}|1^{+}_{m_i} \rangle=
\sum_{pn}
[X_{pn}(1^{+}m_i)X_{pn}(1^{+}m_f)-Y_{pn}(1^{+}m_i)Y_{pn}(1^{+}m_f)]
.\label{overlap}
\end{equation}

\section{Calculation results}

In this section we present the $2\nu\beta\beta$-decay
results obtained within the FR-QRPA for $^{76}$Ge, $^{82}$Se, $^{100}$Mo, $^{116}$Cd, $^{128}$Te  and $^{130}$Xe,
in comparison with the QRPA and SCQRPA ones.
Rather small model bases listed in the Table I are used in order to get full convergence in the FR-QRPA method.
The levels are in a vicinity of the Fermi levels and spin-orbit partners are always taken into account.
FR-QRPA method is rather sensitive to the differences between occupation probabilities
for protons and neutrons entering the denominator in the expression of the bifermionic
operators ${\bar{A}}^{\dagger}, \bar{A}$ (\ref{Qa1}) and in the expresion for $ {\cal R}_{pn}$
factor
of renormalization matrices (\ref{Dpn}).

For levels far from the Fermi one, the
values for occupation probabilities for protons and neutrons become almost equal.
Because of the denominator which appears in the expression of bifermionic operators (\ref{Qa1}),
that causes numerical problems, in particular the method doesn't converge for large enough
values of particle-particle strength.
Therefore, the bases are fixed in order to get convergence for a larger interval
of particle-particle strength, in particular up to the point of the
collapse of the $2\nu\beta\beta$-decay matrix elements.
%The single-particle bases for each nuclei are shown in Table I.

The single particle energies are obtained by using a Coulomb-corrected Woods-Saxon potential
with Bertsch parametrization. The proton and neutron pairing gaps are determined
 phenomenologically to reproduce the odd-even mass differences through a symmetric five-term formula \cite{waps}.
Then the equations for the chemical potentials (\ref{lambdas}) are solved
for proton and neutron subsystems.
The pairing gaps entering the BCS equations are given in the Table I.

The calculation of the QRPA energies and wave functions requires the knowledge of the
particle-hole $\chi$ and particle-particle $\kappa$ strengths of the residual interaction.
The value of particle-hole strength $\chi$ parameter for each nucleus is fixed in order
to reproduce the experimental position of the Gamow-Teller giant resonance in odd-odd
intermediate nucleus as obtained from the (p,n) reactions
\cite{madey}, \cite{ejiri}, \cite{akim}. Those values are also given in the Table I.
The particle-particle strength $\kappa$ is considered as a free parameter.

The numerical results are shown for two groups of nuclei, the nuclei with $A\leq 100$ and
$A>100$ respectively. The calculations are done within QRPA, SCQRPA and FR-QRPA in order to show
the better stability of the latter method.

The main drawback of the QRPA is the overestimation of the ground state correlations
leading to the collapse of the QRPA ground state, near a certain critical interaction
strength. Around this point the backward-going RPA amplitudes $Y_{pn}$ of the first
$1^{+}$ states become overrated, and too many correlations in the ground state are generated
with increasing strength of the particle-particle interaction.
This phenomenon, as a result of the quasiboson approximation used, leads to QRPA collapse
and implies an ambiguous determination of the
$\beta$ and $2\nu\beta\beta$-decay matrix elements.

In Fig.1 and Fig.2  the dependence of the energy of the first excited
Gamow-Teller state in daughter nuclei is plotted versus the $\kappa$ parameter.
Hereafter, the dashed line corresponds to the QRPA case, the dotted line represents the SCQRPA
case and the solid line describes FR-QRPA case. For all studied nuclei the collapse
of the first excited state is shifted to higher values of $\kappa$ for each method
and the stability increases in the FR-QRPA case.
In Fig.3 and Fig.4 the $2\nu\beta\beta$-decay matrix elements as a function of the particle-particle
strength $\kappa$ are shown. The calculations are done for all nuclei within the three
metods. The horizontal dashed line indicates the experimental values taken from \cite{Vog}.

For all nuclei there is a similar behaviour in the sense that QRPA and SCQRPA collapse a bit earlier than the FR-QRPA does.
Although the
chosen bases are rather small, the new effects we intend to emphasise
%by pointing out
as the differences between QRPA extensions are evident.
The FR-QRPA method offers considerably less sensitive
dependence of $M^{2\nu}_{GT}$ on $\kappa$ and shifts the collapse to larger values of particle-
particle strength.

According to the definition (\ref{ISR}) $S^{-}$ ($S^{+}$) is the total summed
Gamow-Teller $\beta^{-} $($\beta^{+}$) transition strength from the ground state of an
even-even nucleus. In Fig.5 and Fig.6 we plot the relative $\beta^{-}$ strength,
$S^{-}/3(N-Z)$ for the mother nucleus (left side) and relative $\beta^{+}$ strength,
$S^{+}/3(N-Z)$ for the daugther (right side), for $A\leq 100$ and $A>100$ respectively as
a function of particle-particle interaction parameter in order to show the magnitude and the nature of violation of ISR.

Now we would like to discuss the conservation of the Ikeda sum
rule $ISR=S_{-}-S_{+}=3(N-Z)$ in the FR-QRPA framework and to compare with the previous
calculations for QRPA ans SCQRPA. We didn't include the calculations for RQRPA because
SCQRPA goes beyond  and brings more improvements than RQRPA, especially for Ikeda sum
rule.

Finally we combine the data of previous plots and show the ratio of SCQRPA and FR-QRPA sum
$ISR/3(N-Z)$ as a function of $\kappa$.
In the QRPA the Ikeda sum rule is exactely conserved as long as all spin-orbit partners of
the single-particle orbitals are included. In the other extended versions of QRPA
the sum rule is violated with a degree of deviation lying between $17\%$ (RQRPA) and
$3\%$ (SCQRPA) \cite{Sto01}. In our study, following the analytical calculation of
\cite{Rod02}, we have shown numerically that Ikeda sum rule is
exactly fullfiled within FR-QRPA formalism.

\section{Conclusions}

%The Fully-Renormalized QRPA (FR-QRPA) is applied in the present paper to calculate the two-neutrino double beta decay amplitudes and
%relevant quantities for $^{76}$Ge, $^{82}$Se, $^{100}$Mo, $^{116}$Cd, $^{128}$Te  and $^{130}$Xe nuclei.
%In spite of the use of rather small model spaces in order to get full convergence in the model,
%the effect of the restoration of the Ikeda sum rule on the decay amplitudes within FR-QRPA has been analyzed in comparison with
%the corresponding results of both QRPA and the self-consistent QRPA.

In summary, the first calculation of the $2\nu\beta\beta$-decay matrix elements within the
recently proposed Fully-Renormalized QRPA, which fulfills Ikeda sum rule exactly, are presented.
The considered nuclear model includes the separable Gamow-Teller residual interaction.
The subject of interest is the effect of the restoration of the Ikeda sum rule on the
$2\nu\beta\beta$-decay observable for $A=76, 82, 100, 116, 128, 130$ systems. Within the present work
we arrived to the folowing important conclusions:\\

i) The SCQRPA violates the Ikeda sum rule. This phenomenon has been indicated in
the previous studies \cite{Sto01}, but the degree of violation we obtained is less than in the
other calculations because we did not include all multipolarities.

ii) In the limit when the difference between proton and neutron quasiparticle occupation numbers
is neglected the FR-QRPA coincides with SCQRPA.

iii) From a comparison of FR-QRPA with SCQRPA results we conclude that the effect of the restoration of the Ikeda sum
rule is important in the range of large value of particle-particle strength beyond the point of collapse of the standard QRPA.

It is worth to mention that the FR-QRPA approach is sensitive to the precise evaluation of the
proton and neutron quasiparticle occupation numbers. Due to the limitation of the approximate
expression given in (\ref{occup}) (motivated by a similarity to the SQRPA approach)
the convergence of the FR-QRPA is achieved only for relatively small model space. However, even
for such a model space the differences among the standard QRPA, SCQRPA, FR-QRPA approach are evident for
$\kappa$ close to the point the standard QRPA breaks up. There is a hope that for a proper ansatz of the
RPA ground state the FR-QRPA approach can work also for a large model space. This is the subject of our
further study.

\acknowledgements

This work was supported in part by the
Landesforschungsschwerpunktsprogramm Baden-Wuerttemberg
"Low Energy Neutrino Physics".
L.P. and V.R. would like to thank the Graduiertenkolleg
"Hadronen im Vakuum, in Kernen und Sternen"
GRK683 and IKYDA02 project for support.
The work of F. \v{S}. was supported in part by the Deutsche
Forschungsgemeinschaft (436 SLK 17/298) and by the
VEGA Grant agency of the Slovac Republic under contract
No. 1/0249/03.

%\newpage
%%%%%%%%%%%%%%%%%%
%% Table I
%%%%%%%%%%%%%%%%%%

%\begin{table}
%\caption{The single particle basis for all the nuclei under consideration}
%\begin{center}
%\begin{tabular}{|c|c|c|}

%&$^{76}$Ge $\rightarrow$ $^{76}$Se & $^{82}$Se $\rightarrow$ $^{82}$Kr \\
%\hline
%Basis &                                          &                                         \\
%      & 1p1/2, 1p3/2, 0f5/2, 0f7/2, 0g7/2, 0g9/2 & 1p1/2, 1p3/2, 0f5/2, 0f7/2, 0g7/2, 0g9/2 \\
%\hline
%\hline

%&$^{100}$Mo $\rightarrow$ $^{100}$Ru & $^{116}$Cd $\rightarrow$ $^{116}$ Sn \\%\hline
%Basis & 1p1/2, 1p3/2                             & 1p1/2, 1p3/2, 0h9/2, 0h11/2,              \\
%      & 2s1/2, 1d3/2, 1d5/2, 0g7/2, 0g9/2        & 2s1/2, 1d3/2, 1d5/2, 0g7/2, 0g9/2          \\
%\hline
%\hline
%& $^{128}$Te $\rightarrow$  $^{128}$Xe & $^{130}$Te $\rightarrow$ $^{130}$ Xe \\
%\hline
%Basis & 0h11/2,0h9/2                             & 0h11/2, 0h9/2                            \\   
%      & 2s1/2, 1d3/2, 1d5/2, 0g7/2, 0g9/2        & 2s1/2, 1d3/2, 1d5/2, 0g7/2, 0g9/2                    
%\end{tabular}
%\end{center}
%\end{table}

\newpage
%%%%%%%%%%%%%%%%%%
%% Table II
%%%%%%%%%%%%%%%%%%

\begin{table}[t]
\caption{ The proton and neutron pairing gaps determined phenomenologically to reproduce 
the odd-even mass difference and the particle-hole strength $\chi$ chosen to reproduce
the experimental position of Gamow-Teller resonance. The single particle basis for all nuclei
under consideration is also shown.}
\begin{center}
\begin{tabular}{|c|cc|cc|cc|}

&$^{76}$Ge &$^{76}$Se & $^{82}$Se & $^{82}$Kr & $^{100}$Mo &$^{100}$ Ru\\
\hline
Basis & 1p, 0f, 0g && 1p, 0f, 0g && 1p, 2s, 1d, 0g& \\
\hline
$\Delta_p$ [MeV] & 1.561 & 1.751 & 1.401 & 1.734 & 1.612 & 1.548\\
$\Delta_n$ [MeV] & 1.535 & 1.710 & 1.544 & 1.644 & 1.358 & 1.296 \\
\hline

$\chi$  [MeV] & 0.21  &       & 0.18  &       & 0.17  &     \\
\hline
\hline
&$^{116}$Cd &$^{116}$Sn & $^{128}$Te & $^{128}$Xe & $^{130}$Te&$^{130}$ Xe\\
\hline
Basis & 1p, 2s, 1d, 0g, 0h && 2s, 1d, 0g, 0h && 2s, 1d, 0g, 0h & \\
\hline
$\Delta_p$ [MeV] & 1.493 & 1.763 & 1.127 & 1.177 & 1.299 &1.043\\
$\Delta_n$ [MeV] & 1.377 & 1.204 & 1.307 & 1.266 & 1.243 &1.180 \\
\hline
$\chi$  [MeV] & 0.14  &       & 0.14  &       & 0.12  &     \\
\end{tabular}
\end{center}
\end{table}

\newpage%%%%%%%%%%
%% Fig.1
%%%%%%%%%%

\begin{figure}
\begin{center}
\begin{tabular}{cc}
{\epsfig{figure=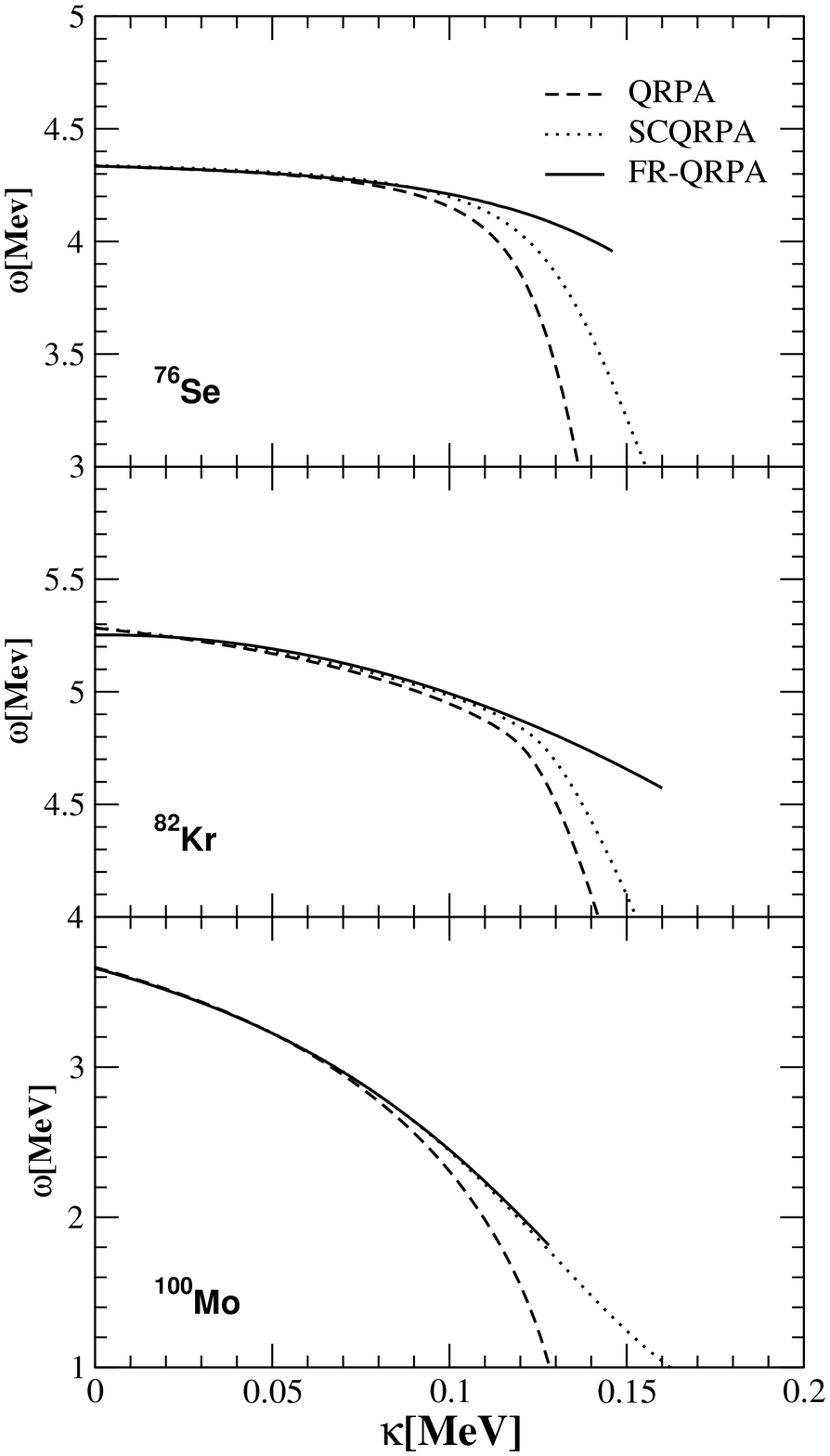,height=18.cm} }
\end{tabular}
\end{center}
\caption{The energy of the first excited Gamow-Teller state as function of particle-particle
interaction strength $\kappa$ for the daughter nuclei with $A \leq 100$}
\label{fig.1}
\end{figure}

%%%%%%%%%%
%% Fig.2
%%%%%%%%%%

\begin{figure}
\begin{center}
\begin{tabular}{cc}
{\epsfig{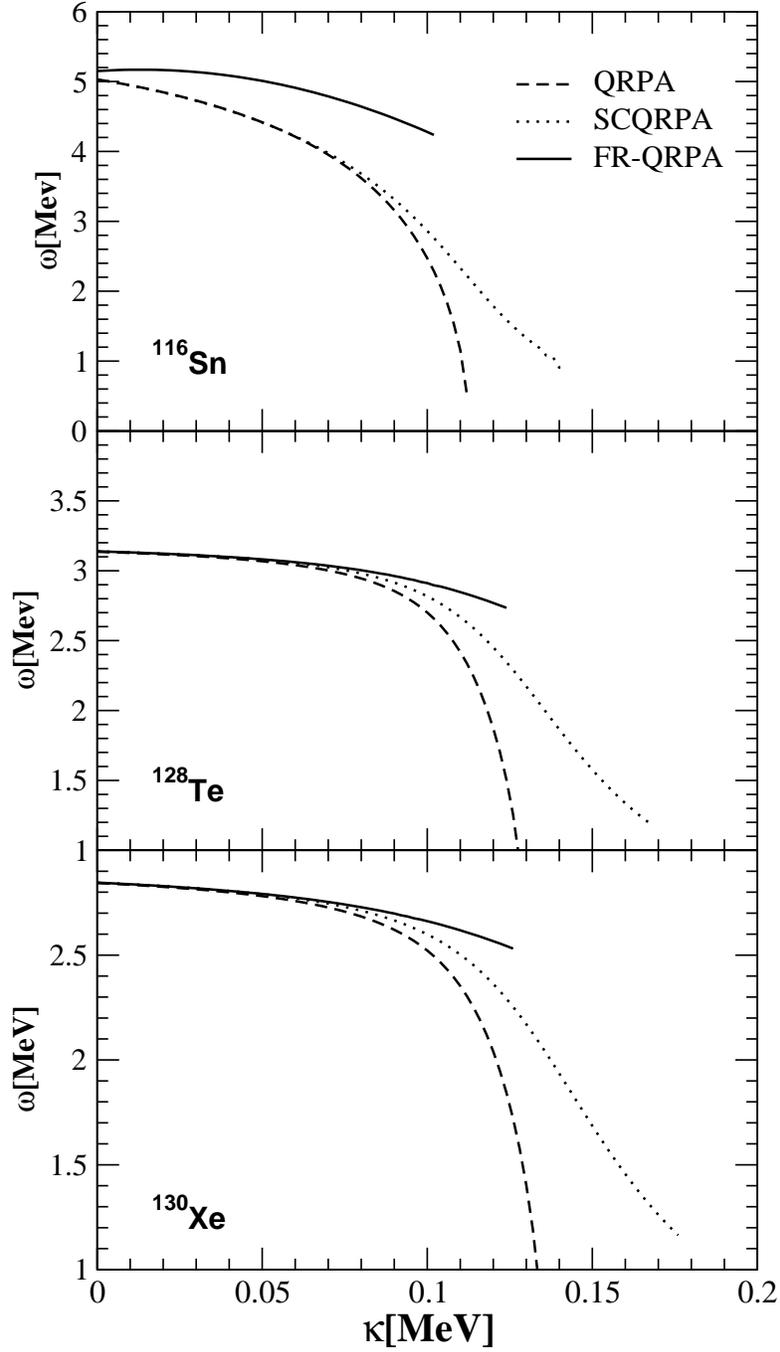} }
\end{tabular}
\end{center}
\caption{The energy of the first excited Gamow-Teller state as function of particle-particle
interaction strength $\kappa$ for the daughter nuclei with $A > 100$}
\label{fig.2}
\end{figure}

%%%%%%%%%%
%% Fig.3
%%%%%%%%%%

\begin{figure}
\begin{center}
\begin{tabular}{cc}
{\epsfig{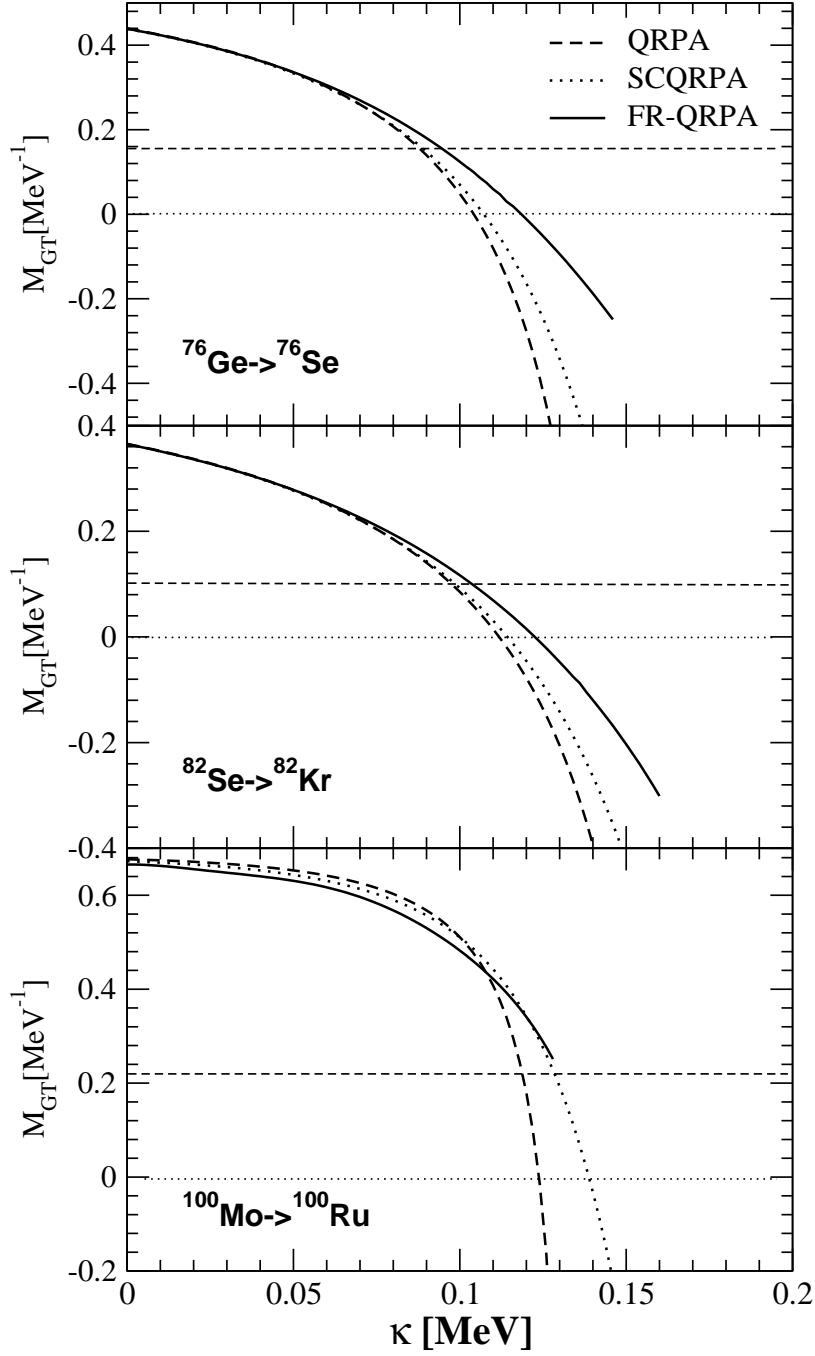} }
\end{tabular}
\end{center}
\caption{$2\nu\beta\beta$-decay matrix elements as a function of particle-particle strength
$\kappa$ for $A \leq 100$ nuclei}
\label{fig.7}
\end{figure}

%%%%%%%%%%
%% Fig.4
%%%%%%%%%%

\begin{figure}
\begin{center}
\begin{tabular}{cc}
{\epsfig{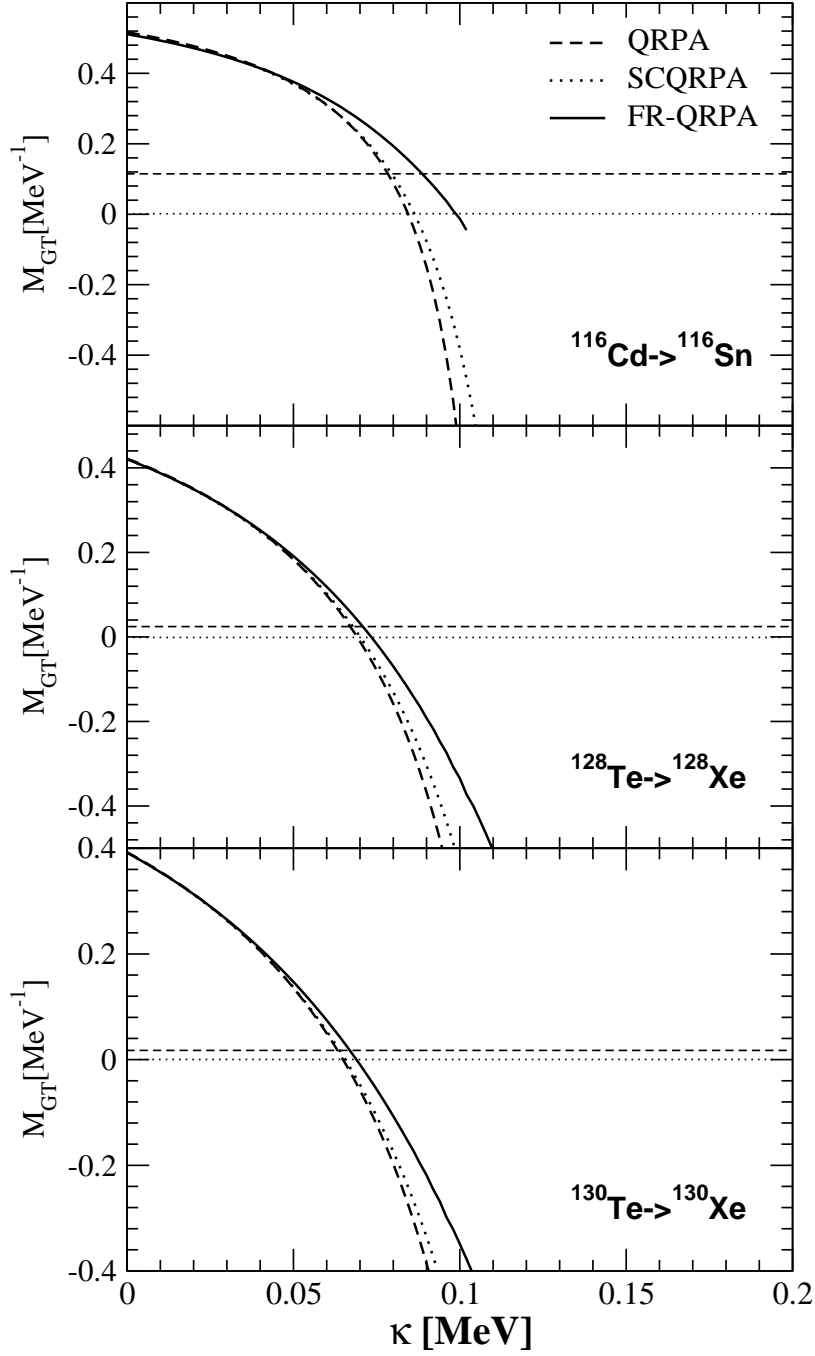} }
\end{tabular}
\end{center}
\caption{$2\nu\beta\beta$-decay matrix elements as a function of particle-particle strength
$\kappa$ for nuclei with $A > 100$}
\label{fig.8}
\end{figure}

%%%%%%%%%%
%% Fig.5
%%%%%%%%%%

\begin{figure}
\begin{center}
\begin{tabular}{cc}
{\epsfig{figure=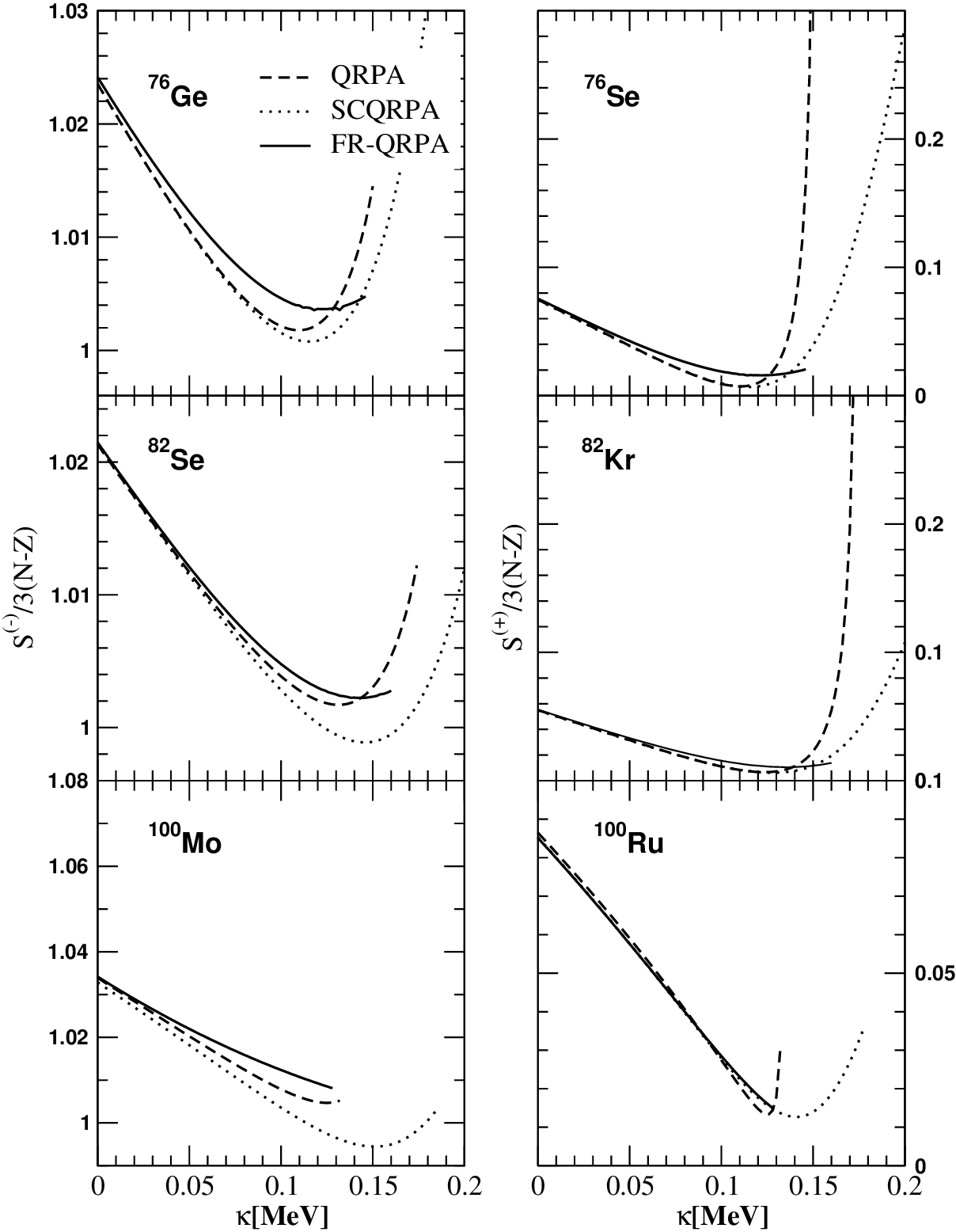,height=18.cm} }
\end{tabular}
\end{center}
\caption{The dependence of the total $\beta^{-}$ and $\beta^{+}$ strength on the $\kappa$ parameter
for mother and daughter nuclei respectively. The total strength is normalized to 3(N-Z) }
\label{fig.3}
\end{figure}

%%%%%%%%%%
%% Fig.6
%%%%%%%%%%

\begin{figure}
\begin{center}
\begin{tabular}{cc}
{\epsfig{figure=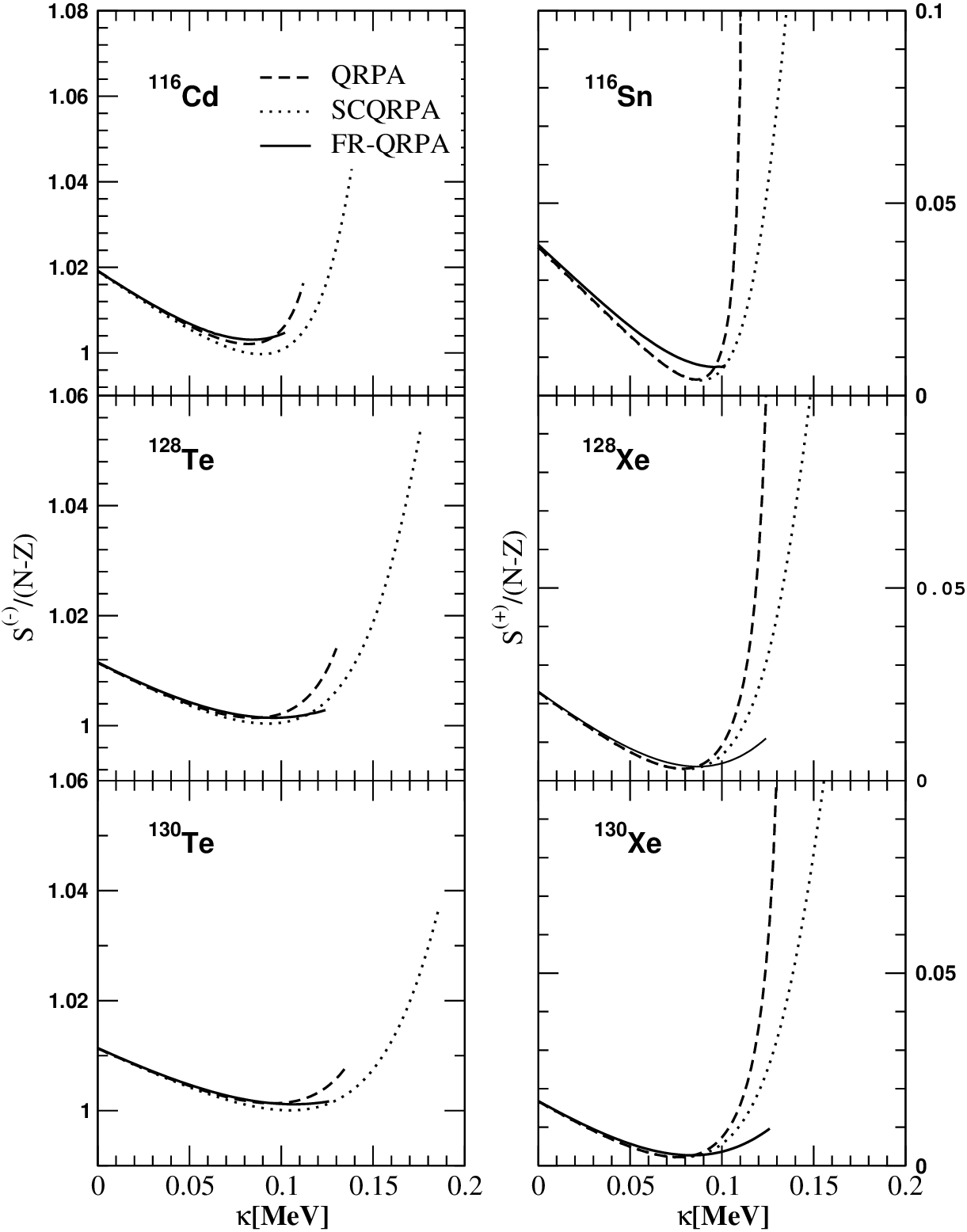,height=18.cm} }
\end{tabular}
\end{center}
\caption{The dependence of the total $\beta^{-}$ and $\beta^{+}$ strength on the $\kappa$ parameter
for mother and daughter nuclei respectively. The total strength is normalized to 3(N-Z)}
\label{fig.4}
\end{figure}

%%%%%%%%%%
%% Fig.7
%%%%%%%%%%

\begin{figure}
\begin{center}
\begin{tabular}{cc}
{\epsfig{figure=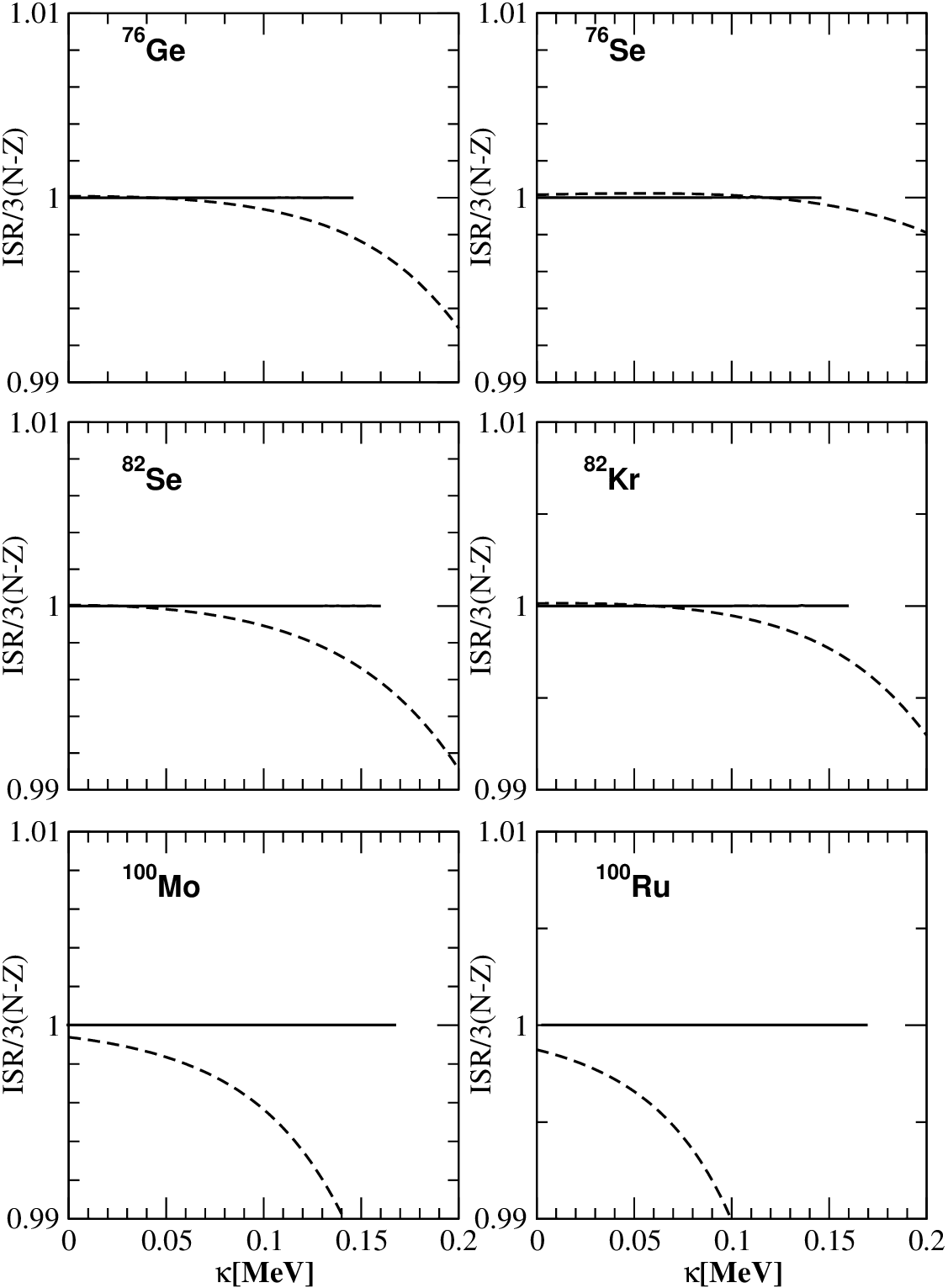,height=18.cm} }
\end{tabular}
\end{center}
\caption{Ikeda sum rule as function of particle-particle interaction strength $\kappa$ 
nuclei with $A \leq 100$. The dashed line coresponds to SCQRPA and the continuous one coresponds 
to FRQRPA.}
\label{fig.5}
\end{figure}

%%%%%%%%%%
%% Fig.8
%%%%%%%%%%

\begin{figure}
\begin{center}
\begin{tabular}{cc}
{\epsfig{figure=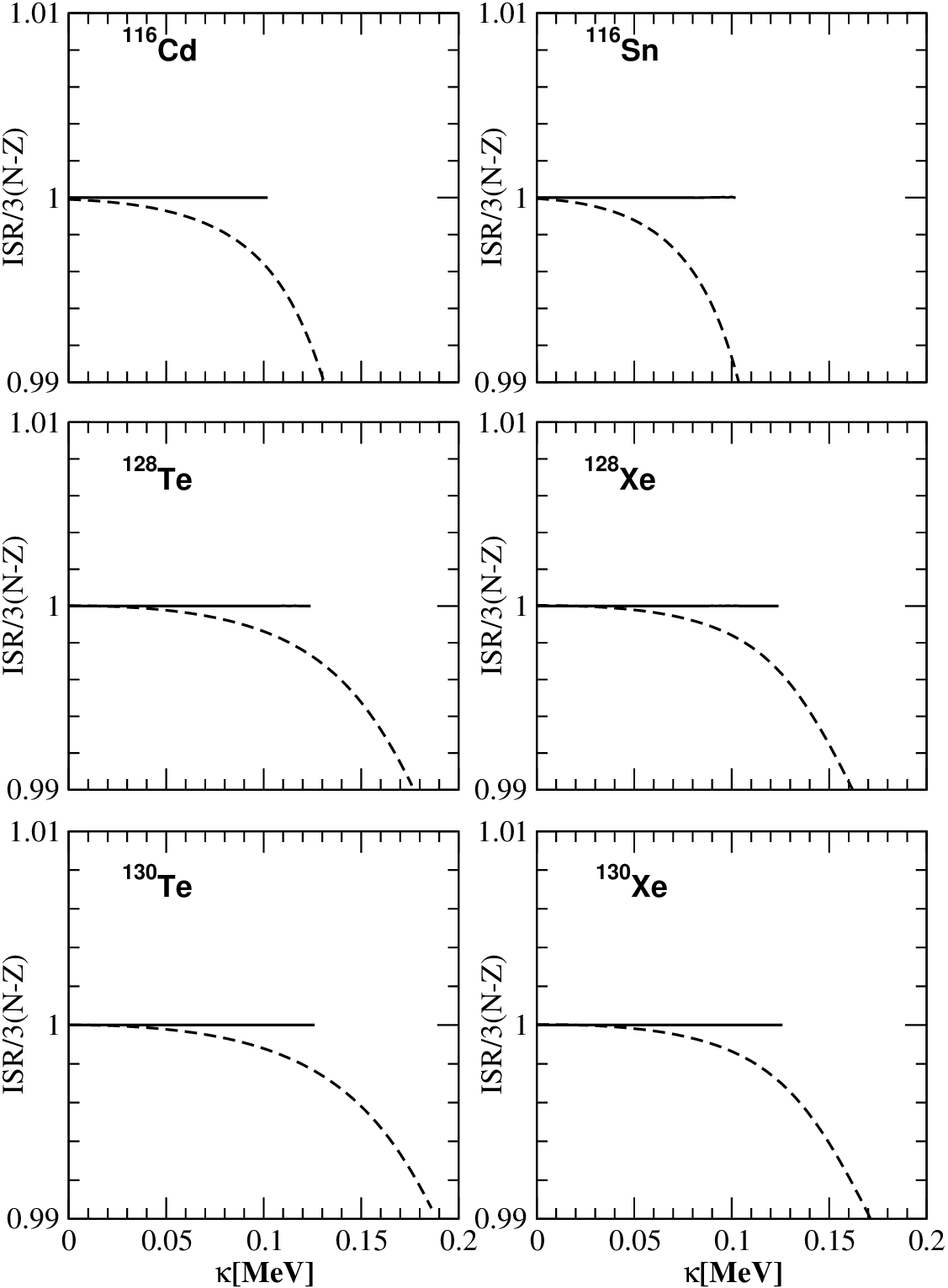,height=18.cm} }
\end{tabular}
\end{center}
\caption{Ikeda sum rule as function of particle-particle interaction strength $\kappa$ 
nuclei with $A < 100$. The dashed line coresponds to SCQRPA and the continuous one coresponds 
to FRQRPA}
\label{fig.6}
\end{figure}

\end{document}